\documentstyle[prl,aps,epsfig,graphics,twocolumn]{revtex}

\def\beq{\begin{equation}}
\def\eeq{\end{equation}}
\def\bea{\begin{eqnarray}}
\def\eea{\end{eqnarray}}

\newcommand{\ket}[1]{| #1 \rangle}
\newcommand{\bra}[1]{\langle #1 |}
\newcommand{\braket}[2]{\langle #1 | #2 \rangle}
\newcommand{\proj}[1]{| #1\rangle\!\langle #1 |}
\newcommand{\ba}{\begin{array}}
\newcommand{\ea}{\end{array}}
\newtheorem{theo}{Theorem}
\newtheorem{defi}{Definition}
\newtheorem{rema}{Remark}
\newtheorem{lem}{Lemma}

\newtheorem{prop}{Property}

\begin{document}

\draft

\title{Separable approximations of density matrices of 
composite quantum systems}

\author{Sini\v{s}a Karnas\cite{poczta1} and Maciej Lewenstein\cite{poczta2}}
\address{Institut f\"ur Theoretische Physik,
Universit\"at Hannover, D-30167 Hannover, Germany
}
\maketitle
\date{\today}

\begin{abstract}
We investigate optimal separable approximations (decompositions) 
of  states $\varrho$ of bipartite quantum systems $A$ and $B$  
of arbitrary dimensions $M\times N$ following the lines of Ref. [M. Lewenstein
and A. Sanpera, Phys. Rev. Lett. {\bf 80}, 2261 (1998)]. Such approximations allow to represent in an optimal way any
 density operator as a sum of a separable state and an entangled state of a certain form. For two qubit systems
($M=N=2$) the best separable  approximation has a form of a mixture of a separable state  and a projector onto a pure
entangled state. We formulate necessary condition that the pure state in the best separable approximation is not to be maximally entangled. We demonstrate that the weight of the entangled state in the best separable
approximation in arbitrary dimensions provides a good entanglement measure.  We prove  in general for arbitrary $M$
and $N$ that
 the best separable approximation  corresponds to a mixture of separable and 
 entangled state both of each are unique. 
We develop also a theory of 
optimal separable approximations for states with positive partial transpose 
(PPT states). Such approximations allow to decompose 
any  density operator with positive
partial transpose as a sum of separable state and an entangled PPT state. We discuss
 procedures of constructing such
decompositions.
\end{abstract}

\narrowtext

\date{\today}


\section{Introduction}

The problem  of characterization of entangled states of composite quantum 
systems is one of the fundamental open problems of quantum theory.  
Entanglement is one of the quantum properties which make quantum mechanics
so fascinating: it leads to famous apparent paradoxes\cite{EPR,Sch}, 
and it is of great importance for applications in 
quantum communication and information processing \cite{effects}.

In the case of the pure  states it is easy to check whether 
a given state is, or is not entangled. So far, the answer to 
this question when applied to quantum mixtures is not known in general.
The definition (introduced by Werner \cite{Werner})
says that a state (in general a mixed state) is entangled when it 
is not separable. Separable states 
defined on a Hilbert space ${\cal H}_{AB}={\cal H}_{A} \otimes 
{\cal H}_{B}$  
are those that can be as a convex combination of projections onto 
product states
\begin{equation} \varrho=\sum_{i=1}^Kp_i |e^{i}_{A},f^{i}_{B}
\rangle \langle e^{i}_{A},f^{i}_{B}|
, \ \ \ \sum_i p_i=1. 
\end{equation}
\vskip0.3cm
In finite dimensional spaces, the number of terms in the sum 
can be restricted to    $K\leq
{\rm dim} ({\cal H}_{AB})^2$ (in another words, when the density 
matrix is separable, then it can be represented in the above form 
with $K$ terms, where $K$ is not larger than the dimension of the space of
linear operators acting in ${\cal H}_{AB}$,
see \cite{tran}).

Several  necessary conditions for separability are known: 
Werner's condition based on the mean value of the, so called,
flipping operator
\cite{Werner}, Horodeckis criterium based 
on  $\alpha$-entropy inequalities
\cite{alfa}, and many others \cite{primer}. Perhaps, the most important necessary criterium 
has been formulated by Peres
\cite{Peres}, who has demonstrated  that the partial transpose 
$\varrho^{T_A}$ of any separable matrix $\varrho$
defined as
$\langle m ,\mu|\varrho^{T_A}| n ,\nu\rangle
= \langle n ,\mu|\varrho| m ,\nu\rangle $
for any fixed orthonormal product basis $|n ,\nu \rangle \equiv 
|e_n \rangle_{A} \otimes |e_\nu \rangle_{B}$
must be positively defined. In the following we will call states  with 
positive partial 
transpose PPT states. Physical meaning of the PPT property is for PPT state
time reversal operation in one
 subsystem (either Alice's or Bob's) is physically sound \cite{primer,tarrach}.

It is worth stressing that the problem of separability is directly  related to
the theory of positive maps on 
$C^*$-algebras\cite{Woronowicz,Stormer} 
This  has been established
in Ref. \cite{sep}, in which it was shown in particular that 
 for systems of low dimensions ($M\times N\le 6$)
the PPT condition is also sufficient for separability. 
For  systems of higher dimensions ($M\times N> 6$) there exist entangled states having 
the PPT property. First examples of such  were provided
 by means of the, so called, range separability criterion
based on analysis of range of density matrix \cite{tran}
(see also \cite{Woronowicz}). Such states represent represent 
bound entanglement, i.e. cannot be distilled \cite{bound}.

In the recent Letter we have also looked at   
the range of the entangled density operators
in order to formulate an  algorithm of optimal decomposition
of mixed states into the separable and inseparable part \cite{M&A}. 
Our method of the {\it best separable approximations} (BSA) 
was based on subtracting projections on product vectors 
from a given density matrix in such a way 
that the remainder remained positively defined. 
This approach allowed to achieve a variety of vary strong results:
optimal decompositions with minimal number of terms  in the form of mixtures and pseudo mixtures 
for $2\times 2$ and $2\times3$
systems\cite{tarrach},
separability criteria for $2\times N$ systems \cite{2xN}, and in general for
$M\times N$ systems (with $M\le N$) \cite{MxN}
for density matrices of low ranks. In particular 
it was shown that: 
i) all PPT states of rank smaller than $N$ are separable; 
ii) for generic states such $r(\varrho)+r(\varrho^{T_A})
\le MN-M-N+2$ constructive separability criteria were 
derived that reduce the problem to finding roots 
of some complex polynomials; iii) for $2\times N$ it  was 
shown that for the states invariant under  partial transpose with respect to the 2
 dimensional subsystem, and those 
that are not ``very different'' from their partial transpose are 
necessarily separable. Very recently, these findings have allowed us to present
general schemes of constructing non decomposable entanglement witnesses (i.e. observables that have a positive mean
value on all separable states, and have a negative mean value on a PPT entangled state\cite{terhal}) and
nondecomposible positive maps 
in arbitrary dimensions, that is the maps that cannot be decomposed into a sum of 
a completely positive map and another completely positive map combined with the transposition
\cite{mapy}. It should be stressed that our approach
goes beyond the methods
of constructing  
examples of PPT entangled states and positive maps based on the, so called. unextendible product
bases\cite{terhal,UPB}.  More importantly, we were able to 
present methods of constructing optimal entanglement witnesses 
and optimal nondecomposible maps which provide very strong separability criteria\cite{opti}. 
In a series of importantant papers Englert and his collaborators have obtained a series of remarkable analytic 
results concerning the BSA decompositions for $2\times 2$ systems \cite{englert}. These results give new deep insight
into the fundamental problem of quantum correlations in 2 qubit systems. 

All of the above mentioned applications 
indicate that the method of BSA is  very useful. 
The aim of this paper is to generalize and to complete
results of the Refs. \cite{M&A}. We present several results that characterize 
the BSA decompositions in 2$\times2$ and, in general in $M\times N$ systems. 
Concerning the 2 qubit systems our results are complementary to those of Ref. \cite{englert}.
The plan of the paper is
as follows: In Section II we remind the reader some basic facts about
the optimal and the best separable approximations. In Section III (using also the
results presented in the Appendix) we demonstrate necessary condition that
for a two qubit systems ($M=N=2$) the best separable 
approximation has a form of a mixture of a separable state 
and a projector to an entangled state which is not {\it maximally entangled}. In Section IV we remind the
reader  the basic facts about entanglement measures; we prove here that
 the weight of the fully entangled 
state in the BSA decomposition of $M\times N$ states  provides a good entanglement
 measure. In Section V we prove 
that  in general for arbitrary $M$ and $N$ 
 the best separable approximation  corresponds to a mixture of separable and 
 entangled state both of each are {\it uniquely} determined. 
Finally, in Section VI we formulate the theory of 
optimal separable approximations for states with positive partial transpose 
(PPT states). Such approximations allow to represent 
any  density operator with positive
partial transpose as a sum of separable state and an entangled PPT state.
Decompositions of this sort play essential role in the theory 
of nondecomposible positive maps \cite{mapy}. We present and discuss efficient numerical
 procedures of construction of such
decompositions.

\section{Introduction to BSA}

Consider a state $\rho$ acting on
${\cal C}^{M} \otimes {\cal C}^{N}$.
Such a state will be called  a PPT state if its partial transpose
satisfies $\rho^{T_{A}}\geq 0$ (or equivalently $\rho^{T_{B}}\geq 0$).
Throughout this paper $K(X), R(X), k(X)$, and $r(X)$ denote the kernel,
the range, the dimension of the kernel, and the rank of the operator $X$,
respectively. By $| e^{*}\rangle$ we will denote the complex conjugated
vector of $| e\rangle$ in the basis $|0\rangle_{A},|1\rangle_{A}, \ldots$ in
which we perform the partial transposition in the Alice space; 
that is, if $| e\rangle
=\alpha| 0\rangle+\beta | 1\rangle+\ldots$ then $| e^{*}\rangle =\alpha^{*}| 0
\rangle+\beta^{*} | 1\rangle+\ldots$. Similar notation will 
be used for vectors in the 
Bob's space.

In this section we give a short repetition of what we call optimal and 
the best 
separability approximations (OSA, and BSA respectively). Although the results 
below have been proven in Ref. \cite{M&A}, we repeat them 
here using the notation of the present work. The idea of BSA 
is that, because of the fact that set of separable states is compact, 
for any density matrix
 $\rho$ there exist a ``optimal" separable matrix $\rho_{s}^{*}$ and "optimal" $\Lambda\ge 0$ such that 
$\Lambda\rho_s^{*}$ can be subtracted 
from $\rho$ maintaining the positivity of the difference,
$\rho-\Lambda\rho_{s}^{*}\geq 0$. This
 situation  is characterized    by the following theorem:

\begin{theo}
For any density matrix $\rho$ (separable, or not) and for any (fixed)
countable set $V$ of product vectors 
belonging to the range of $\rho$, i.e. $|e_{\alpha},f_{\alpha}\rangle
\in R(\rho)$, 
there exist $\Lambda(V)\ge 0$ and  a separable  matrix
\begin{equation}
\rho_{s}^{*}(V)=\sum_{\alpha}\Lambda_{\alpha}P_{\alpha},
\end{equation}
where $P_{\alpha}=|e_{\alpha},f_{\alpha}\rangle\langle e_{\alpha},f_{\alpha}|$,
while  all $\Lambda_{\alpha}\geq 0$, such that 
$\delta\rho=\rho-\Lambda\rho_{s}^{*}\geq0$, and 
that $\rho_{s}^{*}(V)$ provides the optimal separable 
approximation (OSA) to $\rho$  since 
 ${\rm Tr}(\delta\rho)$ is minimal or, equivalently 
$\Lambda$ is maximal. There exists also the best 
separable approximation $\rho_{s}^{*}$ for which  $\Lambda=
{\rm max}_V\Lambda(V)$. Obviously, $\Lambda(V)
\le\Lambda(V')$ when $V'\subset V$
\end{theo}

\begin{rema}
Quite generally one can define the best separable approximations $\rho_s$ of $\rho$ by demanding that
$||\rho-\rho_s||$ is minimal with respect to some norm in the (Banach) space of operators. Here we
minimize ${\rm Tr}(\rho-\lambda\rho_s)$ with respect to all $\rho_s$ such that $\rho-\lambda\rho_s\ge 0$.
\end{rema} 

From this theorem it follows then that if any density
 matrix $\rho$ is separable then $\Lambda=1$. 
Caratheodory's theorem implies then 
(see discussion in Ref. \cite{tran})
that there exist a finite set of 
product vectors $V\subset R(\rho)$ of cardinality $\le r(\rho)^2$, 
for which the optimal separable 
approximation to $\rho$, $\rho_{s}^{*}[V]$ is equal to the BSA and 
$\Lambda=1$ also. The above theorems are also true 
for uncountable families of states $V$, and appropriate generalizations 
are discussed in Ref. \cite{opti}.

In order to explain now how the procedure of construction of the matrix 
$\rho_{s}^{*}$ actually works, we introduce two important concepts:
 
\begin{defi}
A non-negative parameter $\Lambda$ is called {\bf maximal} 
with respect to a (not necessarily normalized) density 
matrix $\rho$, and the projection operator $P=\proj{\psi}$ if  
$\rho-\Lambda P\geq 0$, and for every $\epsilon\geq 0$, 
the matrix $\rho-(\Lambda+\epsilon)P$ is not positive definite.
\end{defi}
This means that $\Lambda$ determines the maximal 
contribution of $P$ that can be subtracted from $\rho$ 
maintaining the non-negativity of the difference. Now we have the 
following important lemma:

\begin{lem}\label{sub}
$\Lambda$ is maximal with respect to $\rho$ and $P=\proj{\psi}$, if: (a) if $\ket{\psi}\not\in R(\rho)$ then
$\Lambda=0$, and (b) if $\ket{\psi}\in R(\rho)$ then
\begin{equation}\label{subeq}
0\leq\Lambda=\frac{1}{\bra{\psi}\rho^{-1}\ket{\psi}}.
\end{equation}
\end{lem}
Note that in the case (b) the expression on RHS of Eq. \ref{subeq}
 makes sense, since $\ket{\psi}\in R(\rho)$, and
therefore  there exists $\ket{\phi}$ such that $\ket{\psi}=\rho\ket{\phi}$, or equivalently that
$\rho^{-1}\ket{\psi}=\ket{\phi}$. Remarkerbly this Lemma has been used in a completely different context by E. Jaynes in his works on foundations of statistical mechanics \cite{jeans}.

\begin{defi}
A pair of non-negative $(\Lambda_{1}$,$\Lambda_{2})$ is 
called {\bf maximal} with respect to $\rho$ and a pair of 
projection operators $P_{1}=\proj{\psi_{1}}$,
$P_{2}=\proj{\psi_{2}}$, if $\rho-\Lambda_{1}P_{1}-\Lambda_{2}P_{2}\geq 0$,
 $\Lambda_{1}$ is maximal with respect to $\rho-\Lambda_{2}P_{2}$ and to the projector $P_{1}$, 
$\Lambda_{2}$ is maximal with respect to $\rho-\Lambda_{1}P_{1}$ and to the projector $P_{2}$, and the sum
$\Lambda_{1}+\Lambda_{2}$ is maximal.
\end{defi}
The condition for the maximality of $\Lambda_1 +\Lambda_2$ is the given by the following lemma:
\begin{lem}\label{parmax}
A pair $(\Lambda_{1},\Lambda_{2})$ is maximal with respect to $\rho$ and 
a pair of projectors $(P_{1},P_{2})$ if: 
\begin{itemize}
\item(a) if $\ket{\psi_{1}}$, $\ket{\psi_{2}}$ do not belong to $R(\rho)$ then 
$\Lambda_{1}=\Lambda_{2}=0$; 

\item (b) if $\ket{\psi_{1}}$ does not belong to $R(\rho)$, while
         $\ket{\psi_{2}}\in R(\rho)$ then $\Lambda_{1}=0$, 
$\Lambda_{2}=\bra{\psi_{2}}\rho^{-1}\ket{\psi_{2}}^{-1}$; 

\item (c) if $\ket{\psi_{1}}$, $\ket{\psi_{2}}\in R(\rho)$ and $\bra{\psi_{1}}\rho^{-1}\ket{\psi_{2}}
=0$,  then $\Lambda_{i}=\bra{\psi_{i}}\rho^{-1}\ket{\psi_{i}}$, $i=1,2$; 

\item (d) if $\ket{\psi_{1}}$, $\ket{\psi_{2}}\in R(\rho)$ 
and $\bra{\psi_{1}}\rho^{-1}\ket{\psi_{1}},\bra{\psi_{2}}\rho^{-1}\ket{\psi_{2}}\ge 
|\bra{\psi_{1}}\rho^{-1}\ket{\psi_{2}}|\not=0$ then
\begin{eqnarray}
\Lambda_{1}=(\bra{\psi_{2}}\rho^{-1}\ket{\psi_{2}}-|\bra{\psi_{1}}\rho^{-1}\ket{\psi_{2}}|)/D,\\
\Lambda_{2}=(\bra{\psi_{1}}\rho^{-1}\ket{\psi_{1}}-|\bra{\psi_{2}}\rho^{-1}\ket{\psi_{1}}|)/D,
\end{eqnarray}
where $D=\bra{\psi_{1}}\rho^{-1}\ket{\psi_{1}}\bra{\psi_{2}}\rho^{-1}\ket{\psi_{2}}-|\bra{\psi_{1}}
\rho^{-1}\ket{\psi_{2}}|^{2}$;

\item (e) finally, if  $\ket{\psi_{1}}$,$\ket{\psi_{2}}\in R(\rho)$ and 
$\bra{\psi_{1}}\rho^{-1}\ket{\psi_{1}}\ge  
|\bra{\psi_{1}}\rho^{-1}\ket{\psi_{2}}|\ge \bra{\psi_{2}}\rho^{-1}\ket{\psi_{2}}$, then $\Lambda_1
=\bra{\psi_{1}}\rho^{-1}\ket{\psi_{1}}^{-1}$, $\Lambda_2=0$.
\end{itemize}
\end{lem}
Note that the Schwarz inequality implies that $D\ge 0$.
We are in the position now to present the the basic BSA theorem:
\begin{theo}
Given the set $V$ of product vectors $\ket{e_{\alpha},f_{\alpha}}\in R(\rho)$, 
the matrix 
$\rho_{s}^{*}=\sum_{\alpha}\Lambda_{\alpha} P_{\alpha}$ is the optimal 
separable approximation (OSA) of $\rho$ if 
\begin{itemize}
\item all 
$\Lambda_{\alpha}$ are maximal with respect 
to $\rho_{\alpha}=\rho-\sum_{\alpha'\not=\alpha}\Lambda_{\alpha'}
P_{\alpha'}$, and to the projector $P_{\alpha}$; 

\item all pairs $(\Lambda_{\alpha},\Lambda_{\beta})$ are 
maximal with respect to $\rho_{\alpha\beta}=\rho-
\sum_{\alpha'\not=\alpha,\beta}\Lambda_{\alpha'}P_{\alpha'}$, 
and to the projection operators $(P_{\alpha},P_{\beta})$.
\end{itemize} 
\end{theo}

If $V$ is 
the set of all product vectors in $R(\rho)$ (in general uncountable) 
then the same theorem holds for the BSA (for the detailed proof see Appendix to Ref. \cite{opti}). All information
about  entanglement is included in the 
matrix $\delta\rho$. If $\delta\rho$ does not vanish, i.e. if 
$\rho$ is not separable, the range $R(\delta\rho)$ cannot contain any 
product vector. The reason is that one can use projectors on
 product vectors that
 belong to $R(\delta\rho)$ in order to increase
 $\Lambda$. The rank of the matrix $\delta\rho$
must be smaller, or equal to  $(M-1)(N-1)$.  This is because the set of all product vectors in 
the Hilbert space $H$ of dimension $M\times N$ spans a $(N+M-1)$-dimensional 
manifold, which generically has a non-vanishing intersection with linear 
subspaces of $H$ of dimension larger than $(N-1)\times (M-1)$.
In fact, we have proven rigorously that this is the case for $2\times N$ systems
in Ref. \cite{2xN}, and presented some rigorous arguments for the case $M 
\times N$ is Ref. \cite{MxN}.  

In particular, for the case of $M=N=2$, 
$\delta\rho$ is a simple projector onto an 
entangled state. For the 2 qubit systems
it is easy to prove that the BSA decomposition 
is unique and has the form:
\begin{equation}\label{ent1}
\rho=\Lambda\rho_{s}+(1-\Lambda)P_{e};\quad\Lambda\in[0,1],
\end{equation} 
where $\rho_{s}$ is the normalized density matrix. If it had not been so,
 we could have another BSA expansion, lets say $\rho=\Lambda\tilde\rho_{s}+
(1-\Lambda)\tilde P_{e}$. But taking the convex combination of these two 
decompositions, we obtain another  BSA decomposition  with the remainder
$\delta\rho$ being given by a convex combination
of $P_{e}$ and $\tilde P_{e}$. Such remainder would have then rank 2, and 
would necessarily contain product vectors in its range \cite{tarrach}. If 
this happened, we would be then able to 
 increase the BSA parameter $\Lambda$ by subtracting from $\delta\rho$ projectors on product vectors in its range. That is, however, 
impossible since $\Lambda$ is already maximal.
For
 the case of arbitrary dimensions the OSA and BSA 
decompositions are also unique. We present 
the proof of this  fact in Section V of this
 paper.

\section{The BSA reminder of ${\cal C}^{2}\otimes {\cal C}^{2}$ quantum systems: is it maximally entangled?}

We have seen that the BSA reminder of ${\cal C}^{2}\otimes {\cal C}^{2}$ quantum systems is just given by a
projector onto a entangled state $\ket{\psi_e}$. This fact is essential and allows to obtain the BSA decomposition for
some states analytically\cite{englert}. For many families of states considered by Englert and his collaborators the BSA remainder
consists of a maximally entangled state. Similar conclusions follow from the numerical analysis of Ref. \cite{M&A}.
In this section we ask therefore a natural question: under which conditions the BSA remainder is, or is not
maximally entangled? Strictly speaking we present here a necessary condition, that the BSA decomposition for a
generic density matrix must fulfill so that the BSA remainder is not maximally entangled.

We concentrate here on generic quantum states  which
have the maximal dimension of the range  $(r(\rho)=r(\rho^{T_{A}})=4)$.  
Let
us assume that the density matrix
$\rho$ has the BSA decomposition 
\begin{equation}
\rho=\Lambda\rho_{s}+(1-\Lambda)P_{\psi_{e}},
\label{disp}
\end{equation}
  so that its partial transposition
with respect to Alice's system,
$\rho^{T_A}=
\Lambda\rho_{s}^{T_A}+(1-\Lambda)P_{\psi_{e}}^{T_A}$. When $\Lambda$ 
is not equal to 1, $\rho$ is entangled, and $\rho^{T_A}$ must 
not be positive definite.

Let us first observe 

\begin{lem}\label{em3}
If $\rho$ acting in ${\cal C}^{2}\otimes {\cal C}^{2}$has the BSA decomposition 
$\rho=\Lambda\rho_{s}+(1-\Lambda)P_{\psi_{e}}$,    then
$r(\rho_s^{T_A})\le 3$.
\end{lem}

\noindent{\bf Proof:}
 Had the
range  of $\rho_s^{T_A}$ been full, one could always replace $1-\Lambda$ by $(1-\Lambda-\epsilon)$, keeping
$\Lambda\rho_{s}^{T_{A}}+\epsilon P^{T_A}_{\psi_{-}}$  positive definite, while $\rho'_s=\rho_s+\epsilon
P_{\psi_{-}}$ separable. $\Box$

The fact that the rank of $\rho^{T_A}$ is not full implies 
that $\exists \ket{v}$, such that $\rho^{T_{A}}\ket{v}=0$. 
Since $P_{\psi}^{T_A}$ 
has $3$ positive and one negative eigenvalue\cite{tarrach}, 
where the eigenvector corresponding to a negative eigenvalue in a conveniently chosen basis 
can be written as 
$$\left(\begin{array}{c}0\\1\\-1\\0\end{array}\right)=\ket{\psi_{-}}$$,
 then $\braket{v}{\psi_{-}}\not= 0$. If it was not the case, one could also
 replace $1-\Lambda$ by $(1-\Lambda-\epsilon)$, keeping $\Lambda\rho_{s}^{T_{A}}+\epsilon P^{T_A}_{\psi_{-}}$ positive.

Let us now discuss the optimization procedure, that sometimes allow
to increase $\Lambda$ in the decomposition (\ref{disp}). A given decomposition of such a form is optimal, if it
cannot be optimized. It will turn out that the optimization strategy works only provided $\psi_e$ is not maximally
entangled. The necessary condition, that the BSA remainder is not maximally entangled, is that the decomposition
cannot be optimized in the sense formulated below. Our aim is to formulate this necessary condition in an
explicit form in this section. 

\noindent{\bf Optimization procedure:} Let us  observe that we can always write 
$$\ket{\psi_{e}}=
N_{1}\ket{e_{1},f_{1}}+N_{2}\ket{e_{2},f_{2}}$$, 
for any basis $\ket{e_{1}}$,$\ket{e_{2}}$, 
where $\braket{e_{1}}{e_{1}}=\braket{e_{2}}{e_{2}}=1$,
 but $\braket{e_{1}}{e_{2}}$ does not have to be zero. 
Let $\ket{\hat e_{1}}$, $\ket{\hat e_{2}}$ denote the basis biorthogonal to 
 $\ket{e_{1}}$, $\ket{e_{2}}$;  we obtain then 
\begin{eqnarray}
\braket{\hat e_{1}}{\psi_{e}}&=&N_{2}\braket{\hat e_{1}}{e_{2}}\ket{f_{2}}\nonumber\\
\braket{\hat e_{2}}{\psi_{e}}&=&N_{1}\braket{\hat e_{2}}{e_{1}}\ket{f_{1}}\nonumber
\end{eqnarray}
Requiring 
that $\braket{f_{1}}{f_{1}}=\braket{f_{2}}{f_{2}}=1$ the above equations allow to 
determine uniquely  $N_{1}$,$N_{2}$,$\ket{f_{1}}$ and $\ket{f_{2}}$. 
Without loosing the generality we may 
assume $N_{1}\geq N_{2}$. 
Let us introduce 
$$\ket{\psi_{e}(\alpha )}=\frac{1}{N(\alpha )}(\alpha N_{1}\ket{e_{1},f_{1}}+\frac{1}{\alpha}N_{2}\ket{e_{2},f_{2}}),$$ 
where $$N(\alpha)^{2}=\alpha^{2} N_{1}^{2}+\frac{1}{\alpha^{2}}N_{2}^{2}+2N_{1}N_{2}Re(\braket{e_{1}}{e_{2}}\braket{f_{1}}{f_{2}}).$$ We can now 
 rewrite the BSA projector
\begin{equation}
P_{\psi_{e}}=N(\alpha)^{2}P_{\psi_{e}(\alpha)}+N_{1}^{2}(1-\alpha^{2})P_{e_{1}f_{1}}+N_{2}^{2}(1-\frac{1}{\alpha^{2}})P_{e_{2}f_{2}}.\label{prj}
\end{equation}
We would like to replace the projector $P_{\psi_{e}}$ by the expression
(\ref{prj}) and in this way improve the BSA decomposition. 
To this aim we  require  that $N(\alpha)^{2}\leq 1$ 
which implies that  $\alpha^{2}N_{1}^{2}+\frac{1}{\alpha^{2}}N_{2}^{2}
\leq N_{1}^{2}+N_{2}^{2}$. 
Defining now $x\equiv\frac{N_{2}^{2}}{N_{1}^{2}}$, 
we see that $N(\alpha)^{2}< 1$ provided 
 $x< \alpha^{2}< 1$. That is only possible if $N_1\ne N_2$. The latter conditions fulfilled if $\psi_e$ is not
maximally entangled, as described in the following lemma:

\begin{lem}\label{lem35}
If $\ket{\psi_{e}}=
N_{1}\ket{e_{1},f_{1}}+N_{2}\ket{e_{2},f_{2}}$,  where $\braket{e_{1}}{e_{1}}=\braket{e_{2}}{e_{2}}=1$, then 
$N_1=N_2$ if $\psi_e$ is maximally entangled.
\end{lem}

\noindent{\bf Proof:} Let us consider a basis in which $\ket{\psi_{e}}=a\ket{00}+\sqrt{1-a^{2}}\ket{11}$,
and assume a general form of  
$\ket{\hat e_{1}}={\sqrt{p}\choose\sqrt{1-p}e^{i\varphi}}$,
$\ket{\hat e_{2}}={\sqrt{p'}\choose\sqrt{1-p'}e^{i\varphi '}}$. In the basis 
considered  we can easy calculate that
\begin{eqnarray}
\braket{\hat e_{1}}{\psi_{e}}=a\sqrt{p}\ket{0}+\sqrt{1-a^{2}}\sqrt{1-p}\ket{1}e^{-i\varphi},\\
\braket{\hat e_{2}}{\psi_{e}}=a\sqrt{p'}\ket{0}+\sqrt{1-a^{2}}\sqrt{1-p'}\ket{1}e^{-i\varphi '},
\end{eqnarray}
so that 
\begin{eqnarray}
N_{2}^{2}|\braket{\hat e_{1}}{e_{2}}|^{2}=a^{2}p+(1-a^{2})(1-p)\\
N_{1}^{2}|\braket{\hat e_{2}}{e_{1}}|^{2}=a^{2}p'+(1-a^{2})(1-p')
\end{eqnarray} 
Note that $|\braket{\hat e_{1}}{e_{2}}|^{2}=|\braket{\hat e_{2}}{e_{1}}|^{2}$, 
so that indeed $N_{1}^{2}=N_{2}^{2}$ if $a^{2}=\frac{1}{2}$, that is when the state $\ket{\psi_{e}}$ is maximally
entangled. $\Box$

Now we can easily prove

\begin{lem}\label{lem4}
If $\rho$ has the BSA decomposition (\ref{disp}), then either $\psi_e$ is maximally entangled, or $r(\rho_s)=3$
\end{lem}

\noindent{\bf Proof:} Suppose that $r(\rho_s)=3$. If $\psi_e$ is not maximally entangled, the optimization procedure
allows to optimize the decomposition by taking 
 $\alpha^{2}< 1$, but very close to one. 
We can indeed improve BSA for $\rho$, 
provided we can subtract $\frac{1-\alpha^{2}}{\alpha^{2}}P_{e_{2}^{*}f_{2}}$ 
from $\Lambda\rho_{s}^{T_{A}}$. This means that $\ket{e_{2}^{*},f_{2}}$ 
must belong to the range $R(\rho_{s}^{T_{A}})$. 
That in turn requires that if $\ket{v}=\ket{\hat e_{1}^{*},h_{1}}+
\ket{\hat e_{2}^{*},h_{2}}$, we then need $\braket{h_{1}}{f_{2}}=0$, or in another words
\begin{equation} 
\braket{v}{e_{2}^{*}}\braket{\hat e_{1}}{\psi_{e}}=0.
\end{equation}
It is easy to see that this equation has many solutions: 
for example take $\ket{e_{2}}=\ket{{\hat e}_{1}}$ and $\ket{{\hat e}_{1}}$ 
proportional to ${1\choose \alpha}=\ket{0}+\alpha\ket{1}$, then the above 
equation implies that 
 $[\braket{v}{0}+\alpha^{*}\braket{v}{1}][\braket{0}{\psi_{e}}+\alpha^{*}\braket{1}{\psi_{e}}]=0$, 
which is a quadratic equation for $\alpha^{*}$ which 
 obviously has solutions for $\ket{e_{2}}\not=\ket{\hat e_{1}}$. 
We conclude that either $r(\rho_{s})=3$, or $N_{1}=N_{2}$. 
The latter can occur if and only if 
 $\ket{\psi_{e}}$ is fully entangled.$\Box$ 
 
Therefore we have to consider the case $r(\rho_{s})=r(\rho_{s}^{t_{A}})=3$.
 From the results presented in the  
Appendix A we know that there exists such a one dimensional family of product states $\ket{e_{2}(\delta),f_{2}(\delta)}$, where $\delta$ is real, such that $\ket{e_{2}(\delta),f_{2}(\delta)}\in R(\rho_{s})$ and $\ket{e_{2}^{*}(\delta),f_{2}(\delta)}\in R(\rho_{s}^{T_{A}})$ is satisfied.

Now we are in the situation where we can explicitly check whether the vector $\ket{\psi_e}$ in the BSA remainder can be
non maximally   entangled. If $\ket{\psi_{e}}$ is given and we have
$\ket{e_{2},f_{2}}=\ket{e(\delta),f(\delta)}$ for a given
$\rho_{s}$, then we can calculate $\ket{f_{1}}$ and $\ket{e_{1}}$ by
\begin{eqnarray}
\ket{f_{1}}&=&\frac{\braket{\hat{e}_{2}}{\psi_{e}}}{|\braket{\hat{e}_{2}}{\psi_{e}}|},\\
\ket{e_{1}}&=&\frac{\braket{\hat{f}_{2}}{\psi_{e}}}{|\braket{\hat{f}_{2}}{\psi_{e}}|},
\end{eqnarray}
and from $\braket{f_{1}}{f_{1}}=1$, we obtain
$|N_{1}|=\frac{|\braket{\hat{e}_{2}}{\psi_{e}}|}{|\braket{\hat{e}_{2}}{e_{1}}|}$. Since we know now
$\ket{e_{1}},\ket{f_{1}}$, we can also easily calculate
$|N_{2}|=\frac{|\braket{\hat{e}_{1}}{\psi_{e}}|}{|\braket{\hat{e}_{1}}{e_{2}}|}$.

We see that the coefficient $N_{1}$ and $N_{2}$ can be explicitly constructed from  $\rho_{s}$ and $\ket{\psi_{e}}$.
We obtain therefore the main result of this section

\begin{theo}
If a generic $(r(\rho)=r(\rho^{T_{A}})=4)$ state $rho$ in ${\cal C}^2\otimes {\cal C}^2$ has the BSA decomposition
$\rho=\Lambda\rho_{s}+(1-\Lambda)P_{\psi_{e}}$, then either $\psi_e$ is maximally entangled, or $r(\rho_s)=
r(\rho_s^{T_A})=3$, and for any
expansion of $\ket{\psi_e}=N_1|e_1,f_1\rangle+N_1|e_2,f_2\rangle$, such that $\ket{e_{2},f_{2}}\in R(\rho_{s})$ and $\ket{e_{2}^{*},f_{2}}\in R(\rho_{s}^{T_{A}})$ holds, it must follow that $N_{1}< N_{2}$.
\end{theo}

\noindent{\bf Proof:} The proof is obvious using the lemmas of this section, and the optimization procedure. If there
exist $\ket{e_{2}(\delta),f_{2}(\delta)}$ such that $N_1>N_2$, the optimization procedure can be applied, which contradicts the optimality of the BSA.
$\Box$

\section{Entanglement measures}

Before we turn to the main results of this paper let us also 
remind the reader in this section some basic facts about entanglement 
measures and their properties.

Once one has  the physical 
picture of entanglement as a resource, one needs
 to formulate this concept mathematically. 
One way leads through a definition of non-entangled, i.e. separable states as discussed in previous sections.
Another possibility is to try to quantify amount of entanglement for a given mixed state. 
The latter approach is realized  by defining
entanglement measures \cite{em1}, and by specifying physical properties
 which the entanglement measure should have. There are several versions of 
definitions of the entanglement measures; here we follow the 
approach of Plenio and Verdal \cite{em2}:

\begin{defi} 
Let $\rho$ be a quantum state acting 
in a Hilbert space ${\cal H}_{AB}={\cal H}_{A} \otimes 
{\cal H}_{B}$, then the function $E(\rho)\mapsto R$ 
is called {\bf{entanglement measure}} if it satisfies:

\begin{enumerate}
\item $E(\rho)=0$, if $\rho$ is separable;

\item Local unitary operation leave $E(\rho)$ invariant, i.e. 
$E(\rho)=E(U_{A}\otimes U_{B}\rho U_{A}^{\dag}\otimes U_{B}^{\dag})$;

\item  Let $\sum_{i}A_{i}A_{i}^{\dag}\otimes B_{i}B_{i}^{\dag}=1$ 
be some complete local measurement (i.e. local positive operator valued map (POVM)), then 
\begin{equation}\label{em3}
E(\rho)\geq\sum_{i}{\rm Tr}(\rho_{i})
E({\rho_{i}}/{{\rm Tr}(\rho_{i})}),
\end{equation}
where $\rho_{i}:=A_{i}\otimes B_{i}\rho A_{i}^{\dag}\otimes B_{i}^{\dag}$. 
This property means that entanglement measure cannot increase in the mean 
under local operations.
\item For pure states the measure of entanglement should reduce
 to the {\bf{entropy of entanglement}}, which is defined as
von Neuman entropy of the reduced density matrix, $\rho_A={\rm Tr}_B\rho$
(or, alternatively $\rho_B={\rm Tr}_A\rho$),
\begin{equation}
E(\rho):=-{\rm Tr}(\rho_{A}\ln\rho_{A});
\end{equation}
\item  Entanglement measure should be {\bf additive} which means that
\begin{equation}
E(\rho_{1}\otimes\rho_{2})=E(\rho_{1})+E(\rho_{2}).
\end{equation} 
\end{enumerate}
\label{meas}
\end{defi}
It should be pointed out that the necessity 
of the last two conditions is still 
disputed in the literature \cite{vidalphd,em3}, 
and therefore we will just concentrate on the first three 
conditions. 
Notice, that in Eq. (\ref{em3}) it may happens that $E({\rho_{i}}/
{{\rm tr}(\rho_{i})})\leq E(\rho)$.

To complete this section, let us list some of the most widely used
entanglement measures. Typically, they fulfill some, but not all  of the conditions 1-5 of the Def. \ref{meas}.
\begin{enumerate}
\item {\bf Entanglement of formation} \cite{em1} is defined as
\begin{equation}
E_{F}:=\min\sum_{i}p_{i}S(\rho_{A}^{i}),
\end{equation}
where $S(\rho_{A}):=-{\rm Tr}(\rho_{A}{\rm ln}\rho_{A})$ is the von 
Neumann entropy and the minimum is taken over all the possible 
realizations of the state, $\rho=\sum_{i}\proj{\psi_{i}}$,
 where $\rho_{A}^{i}={\rm Tr}_{B}(\proj{\psi_{i}})$. 
Notice that in the case where $\rho$ is a pure state ($\rho=\proj{\psi}$), 
the von Neumann entropy of the reduced density matrix
is an entanglement measure. 
The physical meaning of the formation measure is the minimal amount
of pure state entanglement  needed to create a the given entangled state.
Calculation of $E_F$ for a given state is a very difficult task. Remarkably, Wooters, has derived the analytic
formula for $E_F$ for an arbitrary two qubit state \cite{Wooters}.
\item {\bf Relative entropy entanglement measure} \cite{em2} is defined as
\begin{equation} 
E(\rho):=\min_{\rho_{s}}E(\rho||\rho_{s});
\end{equation}
where the minimum is taken over all separable states $\rho_{s}$ and $E(\rho||\rho_{s})$ is the relative entropy, which is given by the expression
\begin{equation}
E(\rho||\rho_{s}):={\rm Tr}(\rho(\ln\rho-\ln\rho_{s}))
\end{equation}

\item {\bf{Bures entanglement measure}} \cite{em1} is defined as
\begin{equation}
E(\rho):=\min_{\rho_{s}}(2-2\sqrt{F(\rho,\rho_{s})}),
\end{equation}
where $F(\rho,\rho_{s})$ is the Uhlmann's fidelity $F(\rho,\rho_{s})
:=({\rm Tr}(\sqrt{\sqrt{\rho}\rho_{s}\sqrt{\rho}}))^{2}$. 
This entanglement measure does not fulfill the last 
two conditions of Definition $3$.
\end{enumerate}

In the recent years a very promising approach has been initiated by Vidal
who has  shown that more 
parameters (the so called entanglement monotones)  are required 
in order to quantify completely 
the non-local character of bipartite pure states \cite{em3}.

\section{The BSA entanglement}
Let us now 
 investigate  how do the local  POVM's
 influence a given BSA decomposition. 
To this aim we consider a POVM of the form of $\sum_{i}V_{i}V_{i}^{\dag}=1,
\quad V_{i}=A_{i}\otimes B_{i}$. After the
$i$-th result is obtained  in the measurement we
obtain  the following density matrix
\begin{eqnarray}
\rho_{i}&:=&\frac{V_{i}\rho V_{i}^{\dag}}{{\rm Tr}(V_{i}\rho V_{i}^{\dag})}\nonumber\\
&=&\Lambda\frac{{\rm Tr}(V_{i}\rho_{s}V_{i}^{\dag})}{{\rm Tr}(V_{i}\rho
V_{i}^{\dag})}\sum_{\alpha}\frac{\Lambda_{\alpha}{\rm Tr}(V_{i}P_{\alpha}V_{i}^{\dag})}{{\rm
Tr}(V_{i}\rho_{s}V_{i}^{\dag})}+\frac{V_{i}P_{\alpha}V_{i}^{\dag}}{{\rm
Tr}(V_{i}P_{\alpha}V_{i}^{\dag})})+\nonumber\\ &+&(1-\Lambda\frac{{\rm Tr}(V_{i}\rho_{s}V_{i}^{\dag})}{{\rm
Tr}(V_{i}\rho V_{i}^{\dag})})(\frac{V_{i}\delta\rho V_{i}^{\dag}}{{\rm Tr}V_{i}\delta\rho V_{i}^{\dag}}).\nonumber
\end{eqnarray}
Defining now
\begin{eqnarray}
\Lambda_{i}&:=&\Lambda\frac{{\rm Tr}(V_{i}\rho_{s}V_{i}^{\dag})}{{\rm Tr}(V_{i}\rho V_{i}^{\dag})},\nonumber\\
\Lambda_{i\alpha}&:=&\Lambda_{\alpha}\frac{{\rm Tr}(V_{i}P_{\alpha}V_{i}^{\dag})}{{\rm
Tr}(V_{i}\rho_{s}V_{i}^{\dag})},\nonumber\\ P_{i\alpha}&:=&\frac{V_{i}P_{\alpha}V_{i}^{\dag}}{{\rm
Tr}(V_{i}P_{\alpha}V_{i}^{\dag})},\nonumber\\
\delta\rho_{i}&:=&\frac{V_{i}\delta\rho V_{i}^{\dag}}{{\rm Tr}(V_{i}\delta\rho V_{i}^{\dag})},\nonumber
\end{eqnarray}
We rewrite the result as:
\begin{displaymath}
V_{i}\rho V_{i}^{\dag}\rightarrow\rho_{i}=\Lambda_{i}\sum_{\alpha}\Lambda_{i\alpha}P_{i\alpha}+(1-\Lambda_{i})\delta\rho_{i}.\label{deco}
\end{displaymath}
We observe that 
\begin{equation}\label{ide}
1-\Lambda=\sum_{i}(1-\Lambda_{i}{\rm Tr}(V_{i}\rho V_{i}^{\dag}))
\end{equation}
holds. Since  for the BSA decomposition of $\rho_{i}$ the inequality
\begin{equation}
\Lambda_{BSA_{i}}\geq\Lambda_{i}
\end{equation}
holds, we get from (\ref{ide}) that
\begin{equation}\label{ep3}
1-\Lambda\geq\sum_{i}(1-\Lambda_{BSA_{i}}{\rm Tr}(V_{i}\rho V_{i}^{\dag})).
\end{equation}
The result (\ref{ep3}) allows to prove the following property:
\begin{prop} The BSA entanglement measure
$$ E(\rho)=1-\Lambda_{BSA}(\rho)$$ fulfills the  
properties 1.--3. of the Def. \ref{meas}.
\end{prop}

{\bf Proof:}

1.   If $\rho$ is separable, i.e. 
$\rho=\rho_{s}$ then $\Lambda=1$, and $E(\rho)=1-\Lambda=0$.

2. If $\tilde\rho=U_{A}\otimes U_{B}\rho U_{A}^{\dag}\otimes U_{B}^{\dag}$ 
then obviously $E(\tilde\rho)  \ge 1-\Lambda=E(\rho)$, and {\it
vice versa}, since we can invert $U_{A}\otimes U_{B}$. 
That means that $E(\rho)$ is invariant with respect to local unitary transformations.

3. Finally, if we apply a local POVM, we obtain
\begin{eqnarray} 
E(\rho)=1-\Lambda&\geq&\sum_{i}(1-\Lambda_{BSA_{i}}
{\rm Tr}(V_{i}\rho V_{i}^{\dag}))\nonumber\\
&\geq&\sum_{i}E(\rho_i){\rm Tr}(V_{i}\rho V_{i}^{\dag}),\nonumber
\end{eqnarray}
where $\rho_i=V_{i}\rho V_{i}^{\dag}/{\rm Tr}(V_{i}\rho V_{i}^{\dag})$.
This follows from (\ref{ep3}).

It is worth noticing  that the above  argument holds for the Hilbert spaces ${\cal H}_A\otimes{\cal H}_B$
of arbitrary dimensions.

\section{The uniqueness of the BSA}

In this Section we turn back to the general case and present a proof that the
BSA in any Hilbert space is unique. To this aim we prove first a lemma, and that 
the major result. 

\begin{lem}\label{mfkt}
Let a  hermitian density matrix $\rho$ has a decomposition of the from
$\rho=\Lambda\rho_{s}+(1-\Lambda)\delta\rho$, where $\rho_{s}$ is the
separable part which has the structure
$\rho_{s}=\Lambda\sum_{\alpha=1}^{n}\Lambda_{\alpha}P_{\alpha}$, with
$P_{\alpha}$ being  the projection operators onto the product states 
$\ket{e_{\alpha}, f_{\alpha}}$ and $\sum_{\alpha=1}^{n}\Lambda_{\alpha}=1$.
Then the set of $\lbrace\Lambda_{\alpha}\rbrace$, which
 are maximal with respect to the density matrix $\rho$ and the set
of the projection operators $\lbrace P_{\alpha}\rbrace$, form a
manifold which generically has a dimension $n-1$ and is determined by the following
equation
\begin{eqnarray}
&1&-\sum_{i}^{n}\Lambda_{i}D_{i}+\sum_{i<j}^{n}\Lambda_{i}\Lambda_{j}D_{ij}-\sum_{i<j<k}^{n}\Lambda_{i}\Lambda_{j}\Lambda_{k}D_{ijk}+\dots\nonumber\\
&+&(-)^{m}\sum_{i_{1}<i_{2}<\dots<\i_{m}}\Lambda_{i_{1}}\Lambda_{i_{2}}\dots\Lambda_{i_{m}}D_{i_{1}i_{2}\dots i_{m}}+\nonumber\\
\dots&+&(-)^{n}\Lambda_{1}\Lambda_{2}\dots\Lambda_{n}D_{12\dots n}=0\label{surf}
\end{eqnarray}
where the set of $\lbrace D_{i_{1}i_{2}\dots i_{m}}\rbrace$ are 
the subdeterminants (minors) of the matrix $D$, which is defined as
\begin{eqnarray}
D=
\left(\begin{array}{cccc}
\bra{\psi_{1}}\rho^{-1}\ket{\psi_{1}}&\bra{\psi_{1}}\rho^{-1}\ket{\psi_{2}}&\dots&\bra{\psi_{1}}\rho^{-1}\ket{\psi_{n}}\\
\bra{\psi_{2}}\rho^{-1}\ket{\psi_{1}}&\bra{\psi_{2}}\rho^{-1}\ket{\psi_{2}}&\dots&\bra{\psi_{2}}\rho^{-1}\ket{\psi_{n}}\\
\vdots&\vdots&\ddots&\vdots\\
\bra{\psi_{n}}\rho^{-1}\ket{\psi_{1}}&\bra{\psi_{n}}\rho^{-1}\ket{\psi_{2}}&\dots&\bra{\psi_{n}}\rho^{-1}\ket{\psi_{n}}\\
\end{array}\right),\nonumber
\end{eqnarray}
and where by  $\lbrace\ket{\psi_{i}}\rbrace$ we denote for shortness 
the product vectors which are building the projection operators $(P_{i}\equiv\proj{\psi_{i}}$.   
\end{lem} 
{\bf Proof: }
 Let us first remark that generically the matrix $D$ does not have a block structure. 
If the matrix $D$ consists  $k$ diagonal $n_k$--dimensional blocks, 
then not only Eq. (\ref{surf}) is fulfilled, but also  the $k$ corresponding equations 
for the blocks, 
so that the corresponding manifold has the dimension $n_k$, and is a cartesian product of 
$k$ manifolds of dimension $n_k-1$. In the following we will concentrate on the generic case. 

The proof of the lemma goes with induction. First we prove it for $n=2$ and we get
\begin{displaymath}
1-\Lambda_{1}D_{1}-\Lambda_{2}D_{2}+\Lambda_{1}\Lambda_{2}D_{12}=0,\nonumber
\end{displaymath} 
or for $n=3$ where we get
\begin{eqnarray}
&1&-\Lambda_{1}D_{1}-\Lambda_{2}D_{2}-\Lambda_{3}D_{3}+
\Lambda_{1}\Lambda_{2}D_{12}+\Lambda_{1}\Lambda_{3}D_{13}+\nonumber\\
&+&\Lambda_{2}\Lambda_{3}D_{23}-\Lambda_{1}\Lambda_{2}\Lambda_{3}D_{123}=0.\nonumber
\end{eqnarray}

Now, let us  assume  that the lemma  is true for $n$, and show that
it must also be true for $n+1$. Let $\rho$ has the decomposition
$\rho=\Lambda\rho_{s}+(1-\Lambda)\delta\rho$, with 
$$\rho_{s}=\Lambda\sum_{\alpha=1}^{n+1}\Lambda_{\alpha}P_{\alpha}.$$
The lemma holds for the matrix $\tilde \rho=\rho-\Lambda_{n+1}\proj{\psi_{n+1}}$
so that the first $n$ coefficient $\Lambda_{\alpha}$ fulfill Eq. (\ref{surf}) with coefficients 
$D$ calculated as above with the substitution $\rho^{-1}\rightarrow \tilde\rho^{-1}=
(\rho-\Lambda_{n+1}\proj{\psi_{n+1}})^{-1}$. 
The latter inverse can be calculated using power series expansion in 
the projector $\Lambda_{n+1}\proj{\psi_{n+1}}$. The result is 
\begin{eqnarray}
(\rho&-&\Lambda_{n+1}\proj{\psi_{n+1}})^{-1}\ket{\psi_{i}}=\rho^{-1}\ket{\psi_{i}}+\nonumber\\
&+&\frac{\Lambda_{n+1}\bra{\psi_{n+1}}\rho^{-1}\ket{\psi_{i}}\bra{\psi_{i}}\rho^{-1}\ket{\psi_{n+1}}}
{1-\Lambda_{n+1}\bra{\psi_{n+1}}\rho^{-1}\ket{\psi_{n+1}}}\rho^{-1}\ket{\psi_{n+1}}.\nonumber
\end{eqnarray}
Inserting the above result to equations defining the surface for the first $n$ 
$\Lambda's$ we get, after tedious, but elementary algebraic calculation
\begin{eqnarray}
&1&-\sum_{i}^{n}\Lambda_{i}D_{i}+\sum_{i<j}^{n}\Lambda_{i}\Lambda_{j}D_{ij}-\sum_{i<j<k}^{n}\Lambda_{i}\Lambda_{j}\Lambda_{k}D_{ijk}+\dots+ \nonumber\\
&+&(-)^{m}\sum_{i_{1}<i_{2}<\dots<\i_{m}}\Lambda_{i_{1}}\Lambda_{i_{2}}\dots\Lambda_{i_{m}}D_{i_{1}i_{2}\dots i_{m}}+\dots\nonumber\\
&+&(-)^{n}\sum_{i_{1}<i_{2}<\dots<\i_{n}}\Lambda_{i_{1}}\Lambda_{i_{2}}\dots\Lambda_{i_{n}}D_{i_{1}i_{2}\dots i_{n}}+\nonumber\\
&+&(-)^{n+1}\Lambda_{1}\Lambda_{2}\dots\Lambda_{n+1}D_{12\dots n+1}=0\nonumber
\end{eqnarray}
which proofs the lemma for $n+1$. $\Box$

Note that in particular, if the decomposition discussed in the above lemma is the BSA, 
then the corresponding $\Lambda's$ fulfill Eq. (\ref{surf}). This observation allows us to 
prove the uniqueness of the BSA in arbitrary dimension. It is important to note that the 
surface  defined by Eq. (\ref{surf}) can be considered for arbitrary $\Lambda$'s, 
not necessarily positive! This surface is 
strictly convex and divides the space of all $\Lambda's$ into two sets:
a convex set of those sets of $\{\Lambda's\}$ which have the property that $\rho-
\Lambda\sum_{\alpha=1}^{n+1}\Lambda_{\alpha}P_{\alpha}$ is positive definite, and concave set
for which the latter matrix is not positive definite. If this surface contains  
a part of a hyperplane (linear subspace), it must contain the whole hyperplane, 
since it is defined by the polynomial equation (\ref{surf}).
This observation is essential to prove the uniqueness of the expansion. 

\begin{lem}[The uniqueness of the BSA]
Any density matrix $\rho$ has a unique decomposition 
$\rho=\Lambda\rho_{s}+(1-\Lambda)\delta\rho$, where $\rho_{s}$ is a 
separable density matrix, $\delta\rho$ is a inseparable matrix with no product 
vectors in its range, and $\Lambda$ is maximal.
\end{lem}

{\bf Proof:} The proof the lemma goes by assuming 
the decomposition is not unique; then there must 
exist at least two BSA decompositions, 
$\rho=\Lambda\rho_{s1}+(1-\Lambda)\delta\rho_{1}$ and $\rho=
\Lambda\rho_{s2}+(1-\Lambda)\delta\rho_{2}$, with the same maximal $\Lambda$. Now, 
 any  convex combination of these two BSA decompositions is also the BSA decomposition,
\begin{eqnarray}
\rho&=&\epsilon\rho_{s1}-(1-\epsilon)\rho_{s2}+\epsilon\delta\rho_{1}+(1-\epsilon)\delta\rho_{2}\nonumber\\
&=&\sum_{i}(\epsilon\Lambda\Lambda_{1i}-(1-\epsilon)\Lambda\Lambda_{2i})P_{i}+(\epsilon\delta\rho_{1}-(1-\epsilon)\delta\rho_{2})\nonumber\\
&\equiv&\rho_{s}(\epsilon)+\delta\rho(\epsilon),\nonumber
\end{eqnarray}
where $\epsilon\in[0,1]$.  The part of the one dimensional hyper plane (line)
$\epsilon\Lambda_{1i}-(1-\epsilon)\Lambda_{2i}$ for  $\epsilon\in[0,1]$
lies on the surface (\ref{surf}).

From the form the surface it follows that the whole line 
$\epsilon\Lambda_{1i}-(1-\epsilon)\Lambda_{2i}$
for all $\epsilon$ lies on that surface. 
This cannot be, since for some $\epsilon\not\in [0,1]$, and  $\delta\rho_{1}\ne\delta\rho_{2}$, 
$\delta\rho(\epsilon)$ must become nonpositive   
definite. This is easy to see since for $\epsilon\to\pm\infty$, $\delta\rho(\epsilon)\propto 
\delta\rho_{1}-\delta\rho_{2}$, and the latter matrix is non zero and has the trace zero, so
that it has to have eigenvalues of opposite signs.
 This is thus a contradiction with 
the assumption made at the beginning, {\it ergo} the BSA decomposition must be unique.

\section{The PPT BSA}

In this section we discuss in detail generalization of the BSA approach for PPT states used in Refs
\begin{theo}
Let $\rho$ be a arbitrary PPT state. For any countable set $V=\{P_{i}=\proj{e_{i},f_{i}}\}$, such that $\ket{e_{i},f_{i}}\in R(\rho)$ and $\ket{e_{i}^{*},f_{i}}\in R(\rho^{T_{A}})$, there exists the best separable approximation of $\rho$ in the form
\begin{equation}
\rho=\Lambda\rho_{s}+(1-\Lambda)\delta\rho,
\end{equation}
where $\rho_{s}=\sum_{i}\Lambda_{i}P_{i}$ is a separable state, $\Lambda$ is maximal, and both $\delta\rho\geq 0$, and $\delta\rho^{T_{A}}\geq 0$. We call such a decomposition a {\bf{PPT BSA}} if it preserves the PPT of the remainder $\delta\rho$ and
\begin{equation}
\Lambda_{PPT}\equiv{\rm{max}_{V}}({\rm{Tr}}(\rho_{s}[V])).
\end{equation}  
\end{theo}

{\bf{Proof:}} Let us consider the set of all separable matrices $\rho_{s}=\sum_{i}\lambda_{i}\proj{e_{i},f_{i}}$, where $\ket{e_{i},f_{i}}\in V$,$\rho-\rho_{s}\geq 0$ and $\rho^{T_{A}}-\rho_{s}^{T_{A}}\geq 0$. This set of $\rho$'s form a convex and bounded set, which means that this set is compact. Because of the compactness there must exist a separable matrix $\rho_{s}$ which has maximal trace $\Lambda={\rm{Tr}}(\rho_{s}[V])$. By expanding $V$ we will finally get the maximal PPT contribution$._{\Box}$

Let us analyze the PPT BSA decomposition in more detail. All information about the PPT entanglement is included in the PPT BSA parameter $\Lambda$ and $\delta\rho$. If the PPT BSA remainder $\delta\rho$ does not vanish, then there exists no product vector $\ket{e,f}$, such that $\ket{e,f}\in R(\delta\rho)$ and simultaneously $\ket{e^{*},f}\in R(\delta\rho^{T_{A}})$ is satisfied. This means that the PPT state $\delta\rho$ is entangled. 

We introduce now, just like in the first version of the BSA, a procedure of constructing the matrix $\rho_{s}$. But before we do this let us define some basic concepts for that: 

\begin{defi}
A non-negative parameter $\Lambda$ is called {\bf{PPT maximal}} with respect to a positive PPT operator $\rho$, and a projection operator $P=\proj{\psi}\in V$ if $\rho-\Lambda P\geq 0$,$\rho^{T_A}-\Lambda\rho^{T_{A}}\geq 0$, and for every $\epsilon\geq 0$, the matrix $\rho-(\Lambda+\epsilon)P$ is not a PPT state.
\end{defi}
This means that the $\Lambda$ is the maximal contribution of $P$ that can be subtracted from $\rho$ by maintaining the PPT of the difference. Now let us introduce the following Lemma:
\begin{lem}
$\Lambda$ is PPT maximal with respect to $\rho$ and $P=\proj{e,f}$ iff:
\begin{itemize}
\item if $\ket{e,f}\not\in R(\rho)$ and $\ket{e^{*},f}\not\in R(\rho^{T_{A}})$, or $\ket{e,f}\not\in R(\rho)$ and  $\ket{e^{*},f}\in R(\rho^{T_{A}})$ or $\ket{e,f}\in R(\rho)$ and $\ket{e^{*},f}\not\in R(\rho^{T_{A}})$ then $\Lambda=0$;

\item if $\ket{e,f}\in R(\rho)$ and $\ket{e^{*},f}\in R(\rho^{T_{A}})$ then
\begin{equation}
\Lambda={\rm{min}}\left((\bra{e,f}\frac{1}{\rho}\ket{e,f})^{-1},(\bra{e^{*},f}\frac{1}{\rho^{T_{A}}}\ket{e^{*},f})^{-1}\right).
\end{equation}
\end{itemize}
\end{lem}
{\bf{Proof: }} From lemma (\ref{sub}) we know that $\Lambda=(\bra{e,f}\frac{1}{\rho}\ket{e,f})^{-1}$ is the maximal contribution to $\rho$ and $\tilde{\Lambda}=(\bra{e^{*},f}\frac{1}{\rho^{T_{A}}}\ket{e^{*},f})^{-1}$ is the maximal contribution to $\rho^{T_{A}}$. In order to maximize and keep the PPT of the difference we have to take the minimum of $\Lambda$ and $\tilde\Lambda$$._{\Box}$

\begin{defi}
A pair of non-negative $(\Lambda_{1},\Lambda_{2})$ is called {\bf{maximal}} with respect to $\rho$ and a pair of projection operators $P_{1}=\proj{e_{1},f_{1}}$ and $P_{2}=\proj{e_{2},f_{2}}$ iff
\begin{itemize}
\item $\rho-\Lambda_{1}P_{1}-\Lambda_{2}P_{2}\geq 0$ and $\left(\rho-\Lambda_{1}P_{1}-\Lambda_{2}P_{2}\right)^{t_{A}}\geq 0$,
\item $\Lambda_{1}$ is PPT maximal with respect to $\rho-\Lambda_{2}P_{2}$,
\item $\Lambda_{2}$ is PPT maximal with respect to $\rho-\Lambda_{1}P_{1}$, and
\item $\Lambda_{1}+\Lambda_{2}$ is maximal.
\end{itemize}
\end{defi}
The conditions for PPT maximizing of pairs $P_{1}=\proj{e_{1},f_{1}}$ and $P_{2}=\proj{e_{2},f_{2}}$ are described in appendix \ref{alg}.

Let us now prove that for a given countable set $V$ of product vectors we can obtain the optimal PPT separable approximation by maximizing all pairs of productvectors in $V$. But before we do this, we have to define the PPT BSA manifold:
\begin{defi}
Let the equation $F(\lambda_{1},\dots,\lambda_{K})=0$ (or $\lambda_{1}=f_{1}(\lambda_{2},\dots,\lambda_{k})$) describes the BSA manifold with respect to $\rho$, and ${\tilde F}(\Lambda_{1},\dots,\lambda_{K})=0$ (or $\lambda_{1}={\tilde f}_{1}(\lambda_{2},\dots,\lambda_{k})$) for $\rho^{t_{A}}$. Without loosing generality in order to obtain the manifold which preserves the PPT of the differenz $(\rho-\rho_{s})$ we have to define
\begin{eqnarray}
\lambda_{1}&=&{\rm{min}}\left(\lambda_{1}=f_{1}(\lambda_{2},\dots,\lambda_{K}),\lambda_{1}=\tilde{f}_{1}(\lambda_{2},\dots,\lambda_{K}\right),\nonumber\\
&\equiv&\bar{f}_{1}(\lambda_{2},\dots,\lambda_{K}).
\end{eqnarray}
The implicit form will then be given by $\bar{F}(\lambda_{1},\dots,\lambda_{K})=0$.
\end{defi}
Notice that the PPT BSA manifold is contineous and all most everywhere differentiable. 

\begin{theo}
Given the set $V$ of product vectors $\ket{e_{i},f_{i}}\in R(\rho)$ where also $\ket{e_{i}^{*},f_{i}}\in R(\rho^{T_{A}})$, then the  matrix $\tilde{\rho}_{s}=\sum_{i=1}\Lambda_{i}P_{i}$ is the optimal PPT separable approximation of $\rho$ if:
\begin{itemize}
\item all $\Lambda_{i}$ are PPT maximal with respect to $\rho_{i}=\rho-\sum_{i'\not=i}\Lambda_{i'}P_{i'}$, and to the projector $P_{i}$;
\item all pairs $(\Lambda_{i},\Lambda_{j})$ are PPT maximal with respect to $\rho_{ij}=\rho-\sum_{i'\not=i,j}\Lambda_{i'}P_{i'}$, and to the projectors $(P_{i},P_{j})$.
\end{itemize}
\end{theo}

{\bf{Proof:}}If $\tilde{\rho}_{s}$ is a PPT BSA decomposition then all $\Lambda_{i}$, as well as all pairs $(\Lambda_{i},\Lambda_{j})$ must be PPT maximal (otherwise maximize $\Lambda_{i}$ would increase the trace of $\tilde{\rho}_{s}$). 

To prove the inverse, consider matrices $\rho_{s}=\sum_{i}\lambda_{i}P_{i}$ for which all individual $\lambda_{i}$ are PPT maximal. This means that $\rho_{s}$ belongs to the boundary of the set $Z$ of all separable matrices such that $\rho-\rho_{s}\geq 0$ and $(\rho-\rho_{s})^{t_{A}}\geq 0$. This boundary is the PPT BSA manifold:
\begin{equation}\label{mfeq}
\bar{F}(\lambda_{1},\dots,\lambda_{K})=0.
\end{equation}
The manifold (\ref{mfeq}) can be written as a function $\lambda_{i}=f_{i}(\{\lambda_{j}\}_{j\not=i})$, depending on which size of the manifold we are. Let $\rho_{s}^{m}=\sum_{i}\Lambda_{i}P_{i}$ be the separable matrix for which all pairs of $\Lambda$'s are PPT maximal. The maximum of $(\Lambda_{i},\Lambda_{j})$ then implies that
\begin{equation}\label{unglei}
\frac{\partial}{\partial\lambda_{i}}\left(\lambda_{i}+f_{j}\right)|_{\lambda=\Lambda}=\frac{\partial}{\partial\lambda_{i}}\left(\sum_{i'\not=j}\lambda_{i'}+f_{j}\right)|_{\lambda=\Lambda}\leq 0,
\end{equation}
for all sides of the manifold $\bar{F}=0$ and $i$,$j$. This means that $\rho_{s}^{m}$ is either a local maximum or a saddle point (not necessary the same derivative in every direction of $\lambda=\Lambda$). Now we have the same situation just like in the original version of the BSA. The later possibility cannot occur, since the set $Z$ is {\bf{convex}} (i.e. if $\rho_{s} ,\rho_{s}'\in Z$ then $\epsilon\rho_{s}+(1-\epsilon)\rho_{s}'\in Z$ for every $0\leq\epsilon\leq 1$). Since \ref{unglei} describes also a convex set it can for sure not be a saddle point. The same argument holds also for the local minimum.
And finally the local maximum must be also a global one, because on a convex set there can not exists two of them. This means that $\tilde{\rho}_{s}=\rho_{s}^{m}$$._{\Box}$

One should stress out at the end of this section, that the PPT BSA can be straight forward generalize to multicomposite systems.

\section{conclusions}

In this paper we have presented several novel results concerning the BSA decmpositions of density matrices of
composite quantum systems. General results concern the uniqueness of the BSA decompositions, the existence of the BSA
entnaglement mass, and the efficient  methods of construction of the BSA decomposition for PPT states.
More specific results for two qubit systems deal with the necessary conditions, that the projector onto a
nonmaximally entnagled state   proviedes the remainder in the BSA decomposition. There are several open questions
concerning the BSA decompositions in higher dimensional Hilbert spaces: what is the structure of remainder in such a
case, how to parametrize the remainders (tha so called edge states \cite{mapy} in the case of PPT BSA).
The physical interpretation of the BSA entanglement mass is not known so far. 
In the case of $2\times 2$ space, we hope that our results, togehter with remarkable analytic results of Englert and
his colleagues\cite{englert} will bring us closer to the challenging goal of analytic construction of the BSA
decomposition for arbitrary two quibit density matrix.

This work has been supported by Deutsche Forschungsgemeinschaft
(SFB 407 and  Schwerpunkt ``Quanteninformationsverarbeitung''), 
and by the IST Programm ``EQUIP''.
We thank D. Bruss, J.I. Cirac, B.-G. Englert, P. Horodecki, B. Kraus, 
 A. Sanpera, R. Werner and M. Wilkens for fruitful
discussions. 
\appendix

\section{Product vectors in the range}
In this appendix we prove some lemmas that has been used in the section IV. 
Both the results as well as the 
proofs are very much parallel to the one used by Woronowicz \cite{Woronowicz}.

\begin{lem}
If $\rho$ is a  density matrix in a $2\times 2$ space   having a positive partial transpose and
$r(\rho)=r(\rho^{T_{A}})=3$,  then there exist a product vector $\ket{e,f}\in R(\rho)$ 
such that $\ket{e^{*},f}\in R(\rho^{T_{A}})$.
\label{lemA}
\end{lem}
{\bf Proof: } Let there be given a density matrix 
$\rho=\left(\begin{array}{cc}A&B\\B^{\dag}&C\end{array}\right)$ 
(A and C are invertible, because otherwise we would have a product vectors 
in the kernel \cite{2xN}, and the existence of $\ket{e,f}$ would follow from the results of Ref. \cite{2xN}). Now, we choose 
the basis in ${\cal H}_{A}$ to $\{\frac{1}{\sqrt{1+|\alpha|^{2}}}{1\choose \alpha},
\frac{1}{\sqrt{1+|\alpha|^{2}}}{-\alpha^{*}\choose 1}\}$. 
In this new basis we obtain that $B(\alpha^{*})=\frac{1}{\sqrt{1+\|\alpha\|^{2}}}
(1\quad -\alpha^{*})\left(\begin{array}{cc}A&B\\B^{\dag}&C\end{array}\right)
{1\choose \alpha^{*}}$ is a function of $\alpha^{*}$ only. 
This means that we can choose $\alpha$ such that ${\rm det}B(\alpha^{*})={\rm det}B^{\dag}(\alpha)=0$. Choosing such an $\alpha$, we get $r(B)=r(B^{*})=1$.\\

The next step is to perform a non unitary, but invertible local transformation 
 $\rho\rightarrow I_{A}\otimes\frac{1}{\sqrt{C}}\rho I_{A}\otimes\frac{1}{\sqrt{C}}$, and 
redefine $A\rightarrow \frac{1}{\sqrt{C}}A\frac{1}{\sqrt{C}}$, $B\rightarrow 
\frac{1}{\sqrt{C}}B\frac{1}{\sqrt{C}}$. After that, the new matrix 
is given by $\rho=\left(\begin{array}{cc}A&B\\B^{\dag}&I\end{array}\right)$. 
Now, we use our assumption that $r(\rho)=3$, from which it follows that $A=BB^{\dag}+\lambda P$, 
where $P$ is a projector on some vector $\ket{\psi}$. The assumption that also $r(\rho^{T_{A}})=3$, 
leads us to $A=B^{\dag}B+\tilde\lambda\tilde P$, where $\tilde P$ is a projector on some other vector $\ket{\tilde\psi}$. This leads us to $BB^{\dag}+\lambda P=B^{\dag}B+\tilde\lambda\tilde P$, 
and since ${\rm tr}(BB^{\dag}-B^{\dag}B)=0$, we get that $\lambda=\tilde\lambda$.\
What is the necessary condition now for ${\ket{f}\choose z\ket{f}}\in r(\rho)$ and 
${\ket{f}\choose z^{*}\ket{f}}\in r(\rho^{T_{A}})$ ? This condition means nothing else than that there 
exist two vectors, lets say ${\ket{h}\choose \ket{g}}$ and ${\ket{\tilde h}\choose \ket{\tilde g}}$, such that
\begin{eqnarray}
\left(\begin{array}{cc}BB^{\dag}+\lambda P&B\\B^{\dag}&I\end{array}\right){\ket{h}\choose \ket{g}}&=&{\ket{f}\choose z\ket{f}},\\
\left(\begin{array}{cc}B^{\dag}B+\lambda\tilde P&B^{\dag}\\B&I\end{array}\right){\ket{\tilde h}\choose \ket{\tilde g}}&=&{\ket{f}\choose z^{*}\ket{f}},
\end{eqnarray}  
from which we get the equation
\begin{equation}\label{g1}
\frac{1}{1-zB}\ket{\psi}=\eta\frac{1}{1-z^{*}B^{\dag}}\ket{\tilde\psi},
\end{equation}
with some complex $\eta$. In order to proof our lemma we must show that there exist a solution for (\ref{g1}). The trick is now to describe the right side of the equation (\ref{g1}) as a complex conjugate of the left side, so that we can construct a solution explicitly.\\
We will show now that the equation $(\ref{g1})$ can indeed be transformed into the form
\begin{equation}\label{g2}
\frac{1}{1-zB}\ket{\psi}=\sigma_{x}\eta\frac{1}{1-z^{*}B^{*}}\ket{\psi^{*}},
\end{equation}
where $\sigma_{x}$ is the Pauli matrix. Defining $\frac{1}{1-zB}\ket{\psi}={v_{1}\choose v_{2}}$, we must have that $v_{1}=\eta e^{i\phi}v_{2}^{*}$ and $v_{2}=\eta e^{i\phi}v_{1}^{*}$. This equation has a solution if $v_{1}=ve^{i\theta}$ and $v_{2}=ve^{i\theta+\delta}$, where $\| v_{1}\| =\|v_{2}\|=v$. Lets take now an arbitrary $\delta$ and require ${1 \choose e^{i\delta}}\sim\frac{1}{1-zB}\ket{\psi}$, which means that
\begin{equation}\label{g3}
\left(\begin{array}{cc}e^{i\delta},&-1\end{array}\right)\frac{1}{1-zB}\ket{\psi}=0
\end{equation}
must hold. Obviously, this equation has not only one solution, but an infinite family of solutions for every $\delta$.

Let us now proof that equation (\ref{g2}) indeed holds. First we choose a  basis $\ket{\psi_{1}}$,$\ket{\psi_{2}}$ such that $B^{\dag}B-BB^{\dag}=\left(\begin{array}{cc}\Lambda & 0\\0 & -\Lambda\end{array}\right)$. Therefore we have that $\lambda(P-\tilde P)=\left(\begin{array}{cc}\Lambda & 0\\0 & -\Lambda\end{array}\right)$. Since the overall phases of $\ket{\psi}$ and $\ket{\tilde\psi}$ are irrelevant, we parameterize $\ket{\psi}$ and $\ket{\tilde\psi}$ in our new basis as $\ket{\psi}={\sqrt{p}\choose\sqrt{1-p}e^{i\phi}}$, $\ket{\tilde\psi}={\sqrt{1-\tilde p}\choose\sqrt{\tilde p}e^{i\tilde\phi}}$. This parameterization yields $\tilde p=p$, $\tilde\phi=\phi$ and $\Lambda=\lambda(1-2p)$.
We observe now that there exist always a unitary $K$ such that $KBK^{\dag}=B^{T}$. From this trivially follows of course that $(K^{\dag})^{T}B^{T}K^{T}=B$, and therefore $(K^{\dag})^{T}KBK^{\dag}K^{T}=B$, from which then $BU=UB$, where $U=K^{\dag}K^{T}$. 

Now we will proof that $K=e^{i\varphi_{0}}\left(\begin{array}{cc}0&1\\1&0\end{array}\right)$.\\
Let $M=BB^{\dag}-B^{\dag}B=\lambda (\tilde P-P)$ (Note that $M=M^{*}$ in our basis). Then we have $KMK^{\dag}=B^{T}B^{*}-B^{*}B^{\dag}=B^{*}(B^{T})^{*}-(B^{\dag})^{*}B^{*}=-M^{*}=-M$. Therefore $M=\lambda (K\proj{\psi}K^{\dag}-K\proj{\tilde\psi}K^{\dag})$, and for the vectors $\ket{\psi}$,$\ket{\tilde\psi}$ we get
\begin{eqnarray}
K\ket{\psi}=\left(\begin{array}{c}e^{i\varphi_{1}}\sqrt{1-p}\\e^{i\varphi_{1}}\sqrt{p}e^{i\phi}\end{array}\right),\nonumber\\
K\ket{\tilde\psi}=\left(\begin{array}{c}e^{i\varphi_{2}}\sqrt{p}\\e^{i\varphi_{2}}\sqrt{1-p}e^{i\phi}\end{array}\right).\nonumber
\end{eqnarray}
This implies $K=\left(\begin{array}{cc}0&e^{i\theta_{1}}\\e^{i\theta_{2}}&0\end{array}\right)$ and therefore $\theta_{2}=\varphi_{1}+\phi$,$\theta_{1}+\phi=\varphi_{1}$,$\varphi_{2}=\theta_{1}+\phi$ and $\varphi_{2}+\phi=\theta_{2}$. But, if $\theta_{1}\not=\theta_{2}$ then $U=\left(\begin{array}{cc}e^{i(\theta_{1}-\theta_{2})}&0\\0&e^{-i(\theta_{1}-\theta_{2})}\end{array}\right)$. $U$ will commute with $B$, if $B$ is diagonal in the chosen basis. But then $BB^{\dag}-B^{\dag}B=0$, from which follows that $\ket{\psi}\sim \ket{\tilde\psi}$, and thus ${\psi \choose 0}$ in the range of $\rho$ which proves the Lemma. This means that $\theta_{1}=\theta_{2}$, and $K=e^{i\varphi_{0}}\sigma_{x}$. Since the overall phases of $K$ are irrelevant, we can assume that $K=\sigma_{x}$. This proves however (\ref{g2}), which consequently proves the Lemma too. 

The reader made think now that we have finished the proof of the Lemma, but remember that at the beginning of the proof we have made a non unitary local operation. What we must do now is to retransform the density matrix $\rho$, and check if our results after that still holds. Let us see what happens after the inverse transformation:
\begin{displaymath}
\rho=\left(\begin{array}{cc}\sqrt{C}BB^{\dag}\sqrt{C}+\lambda \sqrt{C}P\sqrt{C}&\sqrt{C}B\sqrt{C}\\\sqrt{C}B^{\dag}\sqrt{C}&C\end{array}\right)\nonumber
\end{displaymath}
Demanding that ${\ket{f} \choose z\ket{f}}\in R(\rho)$ and ${\ket{f} \choose z^{*}\ket{f}}\in R(\rho^{T_{A}})$ leads to the following conditions: 
\begin{eqnarray}
\frac{1}{1-\sqrt{C}B\frac{1}{\sqrt{C}}z}\sqrt{C}\ket{\psi}&=&\eta\frac{1}{1-\sqrt{C}B^{\dag}\frac{1}{\sqrt{C}}z^{*}}\sqrt{C}\ket{\tilde\psi},\nonumber\\
\sqrt{C}(1-f(z)B)|\psi>&=&\sqrt{C}\eta (1-f^{*}(z)B^{\dag})\ket{\tilde\psi},\nonumber\\
(1-f(z)B)\ket{\psi}&=&\eta (1-f^{*}(z)B^{\dag})\sigma_{x}\ket{\psi}.\nonumber
\end{eqnarray}
We see that the equations are equivalent after the rescaling, so that the Lemma holds .$\qquad\qquad_{\Box}$

The prove of the above Lemma allows to parameterize the set of all product vectors $\ket{e(\delta),f(\delta)}$, which satisfied the condition $\ket{e(\delta),f(\delta)}\in R(\rho_{s})$ and  $\ket{e(\delta)^{*},f(\delta)}\in R(\rho_{s}^{T_{A}})$, by an one dimensional real parameter $\delta$. This will be used in Section III.

\section{PPT pair maximizing}\label{alg}
In this appendix we explain how to PPT maximize a pair of product projectors $(\proj{\psi_{1}}=\proj{e_{1},f_{1}},\proj{\psi_{2}}=\proj{e_{1},f_{1}})$.

As we know from the BSA, the BSA manifold for $\rho$ and $(\proj{\psi_{1}}=\proj{e_{1},f_{1}},\proj{\psi_{2}}=\proj{e_{1},f_{1}})$ is given by
\begin{equation}
F(\Lambda_{1},\Lambda_{2})\equiv 1-\Lambda_{1}D^{0}_{1}-\Lambda_{2}D^{0}_{2}-\Lambda_{1}\Lambda_{2}D^{0}=0,
\end{equation} 
where $D^{0}_{1}=\bra{e_{1},f_{1}}\rho^{-1}\ket{e_{1},f_{1}}$,$D^{0}_{2}=\bra{e_{2},f_{2}}\rho^{-1}\ket{e_{2},f_{2}}$ and $D^{0}=\bra{e_{1},f_{1}}\rho^{-1}\ket{e_{1},f_{1}}\bra{e_{2},f_{2}}\rho^{-1}\ket{e_{2},f_{2}}-\|\bra{e_{1},f_{1}}\rho^{-1}\ket{e_{2},f_{2}}\|^{2}$. But also we have to consider the BSA manifold for $\rho^{T_{A}}$. This one is given by
\begin{equation}
\tilde{F}(\Lambda_{1},\Lambda_{2})\equiv 1-\Lambda_{1}D^{1}_{1}-\Lambda_{2}D^{1}_{2}-\Lambda_{1}\Lambda_{2}D^{1}=0,
\end{equation} 
where $D^{1}_{1}=\bra{e_{1}^{*},f_{1}}(\rho^{t_{A}})^{-1}\ket{e_{1}^{*},f_{1}}$,$D^{1}_{2}=\bra{e_{2}^{*},f_{2}}(\rho^{t_{A}})^{-1}\ket{e_{2}^{*},f_{2}}$ and $D^{1}=\bra{e_{1}^{*},f_{1}}(\rho^{t_{A}})^{-1}\ket{e_{1}^{*},f_{1}}\bra{e_{2}^{*},f_{2}}(\rho^{t_{A}})^{-1}\ket{e_{2}^{*},f_{2}}-\|\bra{e_{1}^{*},f_{1}}(\rho^{t_{A}})^{-1}\ket{e_{2}^{*},f_{2}}\|^{2}$. Now we have to consider two basic cases which can occur.

{\bf{Case 1:}} One of the BSA manifolds is under the other manifold. Without loosing generality we assume that this is $F=0$. Then we have the situation just like in figure $1$.
\begin{figure}
\centerline{\scalebox{0.2}{\epsffile{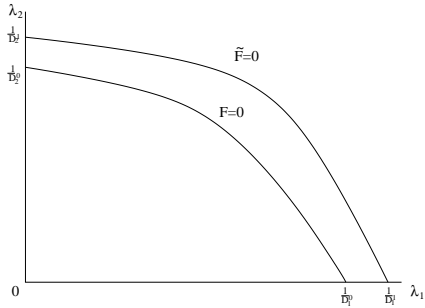}}}
\caption{The Manifold $F=0$ is under $\tilde{F}=0$}
\end{figure}
In that case we have to take the maximum on the manifold $F=0$. From lemma \ref{parmax} we know  the condition for that. Of course we are also including in the case $1$ that there can be an overlap at one endpoints (i.e. if $\frac{1}{D^{0}_{1}}=\frac{1}{D^{1}_{1}}$.

{\bf{Case 2:}} The manifolds have a cross section point between $0<\Lambda_{1}\leq{\rm{max}}\left(\frac{1}{D^{0}_{1}},\frac{1}{D^{1}_{1}}\right)$. Without loosing generality we assume that this describes Figure $2$. Now we can see from Figure $2$ how the PPT BSA manifold $\bar{F}=0$ is constructed, and why it is not differentiable every where. 
\begin{figure}
\centerline{\scalebox{0.2}{\epsffile{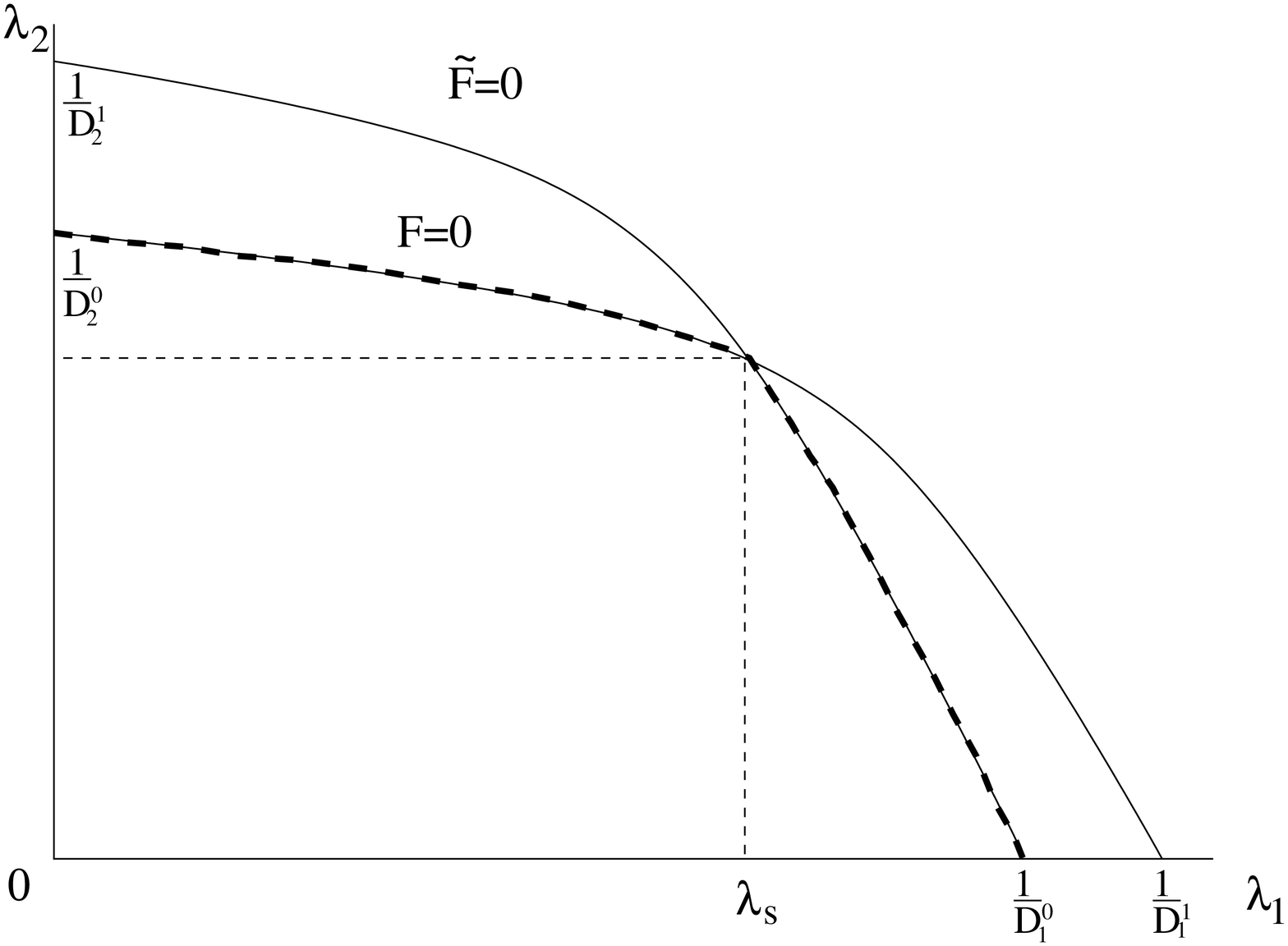}}}
\caption{The manifolds have a cross section point $\lambda_{s}$}
\end{figure}
Let us denote by $\Lambda_{m}$ the maxima of the manifold $F=0$ and also $\tilde{\Lambda}_{m}$ as the maxima of $\tilde{F}=0$. Now we can have the following situations:
\begin{itemize}
\item If $\Lambda_{m}<\lambda_{s}$ and $\tilde{\Lambda}_{m}<\lambda_{s}$ then one has to take $\Lambda_{max}=\Lambda_{m}$;
\item If $\Lambda_{m}>\lambda_{s}$ and $\tilde{\Lambda}_{m}>\lambda_{s}$ then one has to take $\Lambda_{max}=\tilde{\Lambda}_{m}$;
\item If $\tilde{\Lambda}_{m}>\lambda_{s}$ and $\Lambda_{m}<\lambda_{s}$ then one has to take $\Lambda_{max}=\lambda_{s}$;
\item Both maxima are in $\lambda_{s}$, so that $\Lambda_{max}=\lambda_{s}$.
\item The case where $\tilde{\Lambda}_{m}<\lambda_{s}$ and $\Lambda_{m}>\lambda_{s}$ can not occur;
\end{itemize}


\begin{references}
\bibitem[*]{poczta1}
E-mail address: karnas@itp.uni-hannover.de
\bibitem[**]{poczta2}
E-mail address: lewen@itp.uni-hannover.de 

\bibitem{EPR}
A. Einstein, B. Podolsky  and N. Rosen, Phys. Rev.
{\bf 47}, 777 (1935).

\bibitem{Sch}
E. Schr\"odinger, Proc. Cambridge Philos. Soc. {\bf 31},  555 (1935).

\bibitem{effects}
A. Ekert, Phys. Rev. Lett. {\bf 67}, 661 (1991).
C. H. Bennett  and S. J. Wiesner, Phys. Rev. Lett. {\bf 69}, 2881 (1992).
C. Bennett, G. Brassard, C. Crepeau, R. Jozsa, A. Peres and W. K.
Wootters,
Phys. Rev. Lett.  {\bf 70}, 1895  (1993).

\bibitem{Werner}R. Werner,  Phys. Rev. A {\bf 40}, 4277 (1989).


\bibitem{tran}P. Horodecki Phys. Lett. A {\bf 232}, 333 (1997).


\bibitem{alfa} R. Horodecki, P. Horodecki, and M. Horodecki,
Phys. Lett. A {\bf 230}, 377 (1996).

\bibitem{primer} for a revue see M. Horodecki, P. Horodecki and R. Horodecki in ``Quantum Information - Basic Concepts and Experiments'', Eds. G. Alber and M. Weiner, in print (Springer, Berlin, 2000). For a primer see M. Lewenstein, D. Bru{\ss}, J. I. Cirac, B. Kraus, J. Samsonowicz, A. Sanpera and R. Tarrach, quant-ph/0006064.

\bibitem{Peres} A. Peres  Phys. Rev. Lett. {\bf 77}, 1413 (1996).

\bibitem{tarrach} A. Sanpera, R. Tarrach, and G. Vidal, Phys. Rev. A{\bf 58},
826 (1998).


\bibitem{Woronowicz} S. L. Woronowicz, Rep. Math. Phys.,
{\bf 10}, 165 (1976); see also
E. Str\"omer, Acta Math. {\bf 110}, 233 (1963), M. D. Choi, Lin. Alg. and Its. Appl. {\bf 10}, 285 (1975) and M. D. Choi, Proc. Sympos. Pure. Math. {\bf 38}, 583 (1982).

\bibitem{Stormer}S. L. Woronowicz, Commun. Math. Phys. {\bf 51}, 243 (1976); P. Kruszy\'nski and S. L. Woronowicz Lett. Math. Phys. {\bf 3}, 319 (1979). 

\bibitem{sep} M. Horodecki, P. Horodecki, R. Horodecki, 
Phys. Lett. A {\bf 223}, 1 (1996).

\bibitem{bound}
M. Horodecki, P. Horodecki and R. Horodecki, Phys. Rev. Lett.
{\bf 80}, 5239 (1998).

\bibitem{M&A} M. Lewenstein and A. Sanpera, Phys. Rev. Lett. {\bf 80}, 2261
(1998).

\bibitem{2xN} B. Kraus, J. I. Cirac, S. Karnas and M. Lewenstein, Phys. Rev. A{\bf 61}, 062302 (2000).


\bibitem{MxN}  P. Horodecki, M. Lewenstein, G. Vidal and I. Cirac, Phys. Rev. A {\bf 62}, 032310 (2000).

\bibitem{terhal}B. Terhal, quant-ph/9810091; M. Lewenstein, B. Kraus, P. Horodecki and J. I. Cirac, in print in Phys. Rev. A,  quant-ph/0005112.

\bibitem{mapy}  M. Lewenstein, B. Kraus, J. I. Cirac and P. Horodecki, Phys. Rev. A{\bf 62}, 052310 (2000).

\bibitem{UPB}
C. H. Bennett, D. P. DiVincenzo, T. Mor, P. W. Shor, J. A. Smolin, and
B. M. Terhal, Phys. Rev. Lett. {\bf 83}, 3081 (1999); D. P. DiVincenzo,
T. Mor, P. W. Shor, J. A. Smolin and B. M. Terhal, quant-ph/9908070;
C. H. Bennett, D. P. DiVincenzo, Ch. A. Fuchs, T. Mor, E. Rains, P. W. Shor,
J. A. Smolin and W. K. Wootters, quant-ph/9804053; see also
R. Horodecki, M. Horodecki, and P. Horodecki, quant-ph/9811004.

\bibitem{opti} J. I. Cirac, W. D\"ur, B. Kraus and M. Lewenstein, quant-phy/0007057.

\bibitem{englert} B. G. Englert and N. Metwally, J. Mod. Opt. {\bf 47},2221 (2000); B. G Englert and N. Metwally, quant-phy/0007053.

\bibitem{jeans}E. T. Jaynes Phys. Rev. {\bf 106}, 620 (1957); E. T. Jaynes Phys. Rev. {\bf 108}, 171 (1957).

\bibitem{em1}
C. H. Bennett, D. P. DiVincenzo, J. A. Smolin and W. K. Wootters, Phys. Rev. A {\bf 54}, 3824 (1996).

\bibitem{em2}
V. Vedral and  M.B. Plenio,
Phys. Rev. A{\bf 57}, 3 (1998).

\bibitem{vidalphd} G. Vidal, Phys. Rev. A{\bf 59}, 141 (1999). 

\bibitem{em3} G.  Vidal, J. Mod. Opt. {\bf 47}, 355 (2000).

\bibitem{Wooters} W. K. Wootters, Phys. Rev. Lett. {\bf 80}, 2245 (1998).

 
\end{references}
\end{document}